\documentclass{PoS}
\usepackage{multicol}
\usepackage{amsmath}
\def\bea {\begin{eqnarray}}
\def\eea {\end{eqnarray}}
\def\bra{\langle}
\def\ket{\rangle}
\def\nn{\nonumber}

\def\rd{R_D}
\def\rds{R_{D^\ast}}
\def\fbad{\mathcal{A}_{FB}^{D}}
\def\fbads{\mathcal{A}_{FB}^{D^\ast}}
\def\hedt{\bar c \, \sigma^{\mu\nu} \,  b}
\title{A Closer Look at $R_D$ and $R_{D^\ast}$}

\ShortTitle{Looking at $R_D$ and $R_{D^\ast}$}

\author{\speaker{Debjyoti Bardhan}\thanks{The speaker wishes to thank the co-authors on the paper for helping with the talk. Further, thanks is due to Prof. Amol Dighe and Prof. Gautam Bhattacharyya for valuable suggestions before the talk.}\\
        Tata Institute of Fundamental Research\\
        E-mail: \email{debjyoti@theory.tifr.res.in \\
        TIFR/TH/17-13}}

\abstract{The measurement of $R_D$ ($R_{D^*}$), the ratio of the branching fraction of $\overline{B} \to D \tau \bar{\nu}_\tau (\overline{B} \to D^* \tau \bar{\nu}_\tau)$ to that of 
$\overline{B} \to D l \bar{\nu}_l (\overline{B} \to D^* l \bar{\nu}_l)$, shows $1.9 \sigma$ $(3.3 \sigma)$ deviation from 
its Standard Model (SM) prediction. The 
combined deviation is at the level of $4 \sigma$ according to the Heavy Flavour Averaging Group (HFAG). 
We perform an effective field theory analysis (at the dimension 6 level) of these potential New Physics 
(NP) signals assuming $ SU(3)_{C} \times SU(2)_{L} \times U(1)_{Y}$ gauge invariance. We first show that, in general, 
$R_D$  and $R_{D^*}$ are theoretically independent observables and hence, their theoretical predictions 
are not correlated. We identify the operators that can explain the  experimental measurements of $R_D$  and 
$R_{D^*}$ individually and also together. Motivated by the recent measurement of the 
$\tau$ polarisation in $\overline{B} \to D^* \tau \bar{\nu}_\tau$ decay, $P_\tau^{D^*}$ by the {\sc Belle} collaboration, we study the 
impact of a more precise measurement of  $P_\tau^{D^*}$ (and a measurement of $P_\tau^D$) on the various possible NP 
explanations. Furthermore, we show that the measurement of $R_{D^*}$ in bins of $q^2$, the square of the invariant mass of 
the lepton neutrino system, along with the information on $\tau$ polarisation,  can completely distinguish the various 
operator structures.}

\FullConference{9th International Workshop on the CKM Unitarity Triangle\\
		28  November - 3 December 2016\\
		Tata Institute for Fundamental Research (TIFR), Mumbai, India}

\begin{document}

\section{Introduction}
The quantity $R_{D^{(*)}}$ is defined as the following ratio between two branching ratios:
\begin{equation}
R_{D^{(*)}} = \frac{\mathcal{B}(B \to D^{(*)} \tau \bar{\nu}_\tau )}{\mathcal{B}(B \to D^{(*)} l \bar{\nu}_l)}
\end{equation}
where $l = e, \mu$. This quantity, being a ratio, is a `clean' observable devoid of the 
systematic uncertainties that plague individual measurements of branching ratios. Experimental measurements of these 
two quantities - $0.397 \pm 0.028$ for $\rd$ and $0.316 \pm 0.019$ for $\rds$ \cite{Amhis:2014hma} - don't match with 
the theoretical predictions from the Standard Model (SM) - $0.300 \pm 0.011$ for $\rd$ and $0.254 \pm 0.004$ for $\rds$.
This corresponds to deviations $1.9\,\sigma$ and 
$3.3\, \sigma$ significance for $R_D$ and $\rds$ respectively, while the discrepancy for the
two taken together is quite large $\sim 4\, \sigma$ \footnote{It is worth noting that even though the quoted results suggest a large deviation,
a recent measurement of $\rds$ by the {\sc Belle} collaboration \cite{Abdesselam:2016xqt} is consistent with the SM value, although the measurement is quite imprecise.}. This might well be a signal for new physics and we perform a model-independent
analysis of the process using six-dimensional operators; in this analysis,
we assume that any NP only affects the third leptonic generation. 

Besides $\rd$ and $\rds$, we also consider the binned value of $\rd$ and $\rds$, the polarisation of
the final state $\tau$ lepton, $P_\tau^D$ and $P_\tau^{D^\ast}$ and the forward-backward asymmetry in the two processes, $\fbad$ and $\fbads$\footnote{In principle, various differential distributions are also sensitive to the 
different NP Lorentz structures, see e.g., \cite{Datta:2012qk}.}. 
While a recent measurement of $P_\tau^{D^\ast}$ has been reported by {\sc Belle} for the first time (although with large errors) \cite{Abdesselam:2016xqt}, none of the other quantities have been experimentally measured as yet. The definitions of the observables is given below:
\begin{eqnarray}
{\rm Binned} \ \  R_{D^{(*)}} &:& \  \  R_{D^{(*)}} [q^2 {\rm \ \ bin}] = \frac{\mathcal{B}(B \to D^{(*)} \tau \bar{\nu}_\tau )[q^2 {\rm \ \ bin}]}{\mathcal{B}(B \to D^{(*)} l \bar{\nu}_l)[q^2 {\rm \ \ bin}]}\\
{\rm Tau \ \ Polarisation \ \ } &:& \ \ P_\tau^{D^{(*)}}  = {\Gamma^{D^{(*)}}_\tau (+) - \Gamma^{D^{(*)}}_\tau (-) \over \Gamma^{D^{(*)}}_\tau (+) + \Gamma^{D^{(*)}}_\tau (-)}\\ 
{\rm FB \ \ Asymmetry} &:& \mathcal{A}_{FB}^{D^{(\ast)}} = {\int_0^{\pi/2} \frac{d\Gamma^{D^{(*)}}_\tau}{d\theta} d\theta - \int_{\pi/2}^{\pi} \frac{d\Gamma^{D^{(*)}}_\tau}{d\theta} d\theta \over 
\int_0^{\pi/2} \frac{d\Gamma^{D^{(*)}}_\tau}{d\theta} d\theta - \int_{\pi/2}^{\pi} \frac{d\Gamma^{D^{(*)}}_\tau}{d\theta} d\theta} 
\end{eqnarray}

The branching ratio can be written as 
\begin{eqnarray}
{d^2 \mathcal{B}_\ell^{D^{(*)}} \over dq^2 d(\cos \theta)} &=& \mathcal{N} |p_{D^{(*)}}| \left(a_\ell^{D^{(*)}}  + b_\ell^{D^{(*)}} \cos \theta + c_\ell^{D^{(*)}} \cos^2\theta \right)
\label{eqn:br}
\end{eqnarray}
where \[\mathcal{N} = \frac{\tau_B \, G_F^2 |V_{cb}|^2q^2}{256 \pi^3 M_B^2}  \, \left( 1 - \frac{m_\ell^2}{q^2} \right)^2 {\rm \ \ \ and \ \ \  } |p_{D^{(*)}}| = \frac{\sqrt{\lambda(M_B^2, M_{D^{(*)}}^2,q^2)}}{2 M_B}\] where $\lambda(a,b,c)= a^2 + b^2 +c^2 -2 (ab + bc + ca)$ and $\theta$ is 
the angle between the lepton and $D^{(*)}$-meson in the lepton-neutrino centre-of-mass frame.

The decay amplitude for the process can be factorised into two parts - the hadronic part an the leptonic part. The hadronic part of the decay amplitude cannot be calculated exactly and is parameterised using form factors. These form factors are calculated in some 
theoretical and numerical framework and, in this work, we choose to simply borrow those
results. 
\section{Lagrangian and Operator Basis}
The effective six-dimensional Lagrangian for $b \to c \, \ell \, \bar{\nu}_\ell$ we use for the analysis is given by:
\vspace{-3mm}
\begin{multicols}{2}
\noindent
\begin{align}
\label{b2c-basis}
{\cal O}^{cb \ell}_{\rm VL}      &=    [\bar{c} \, \gamma^\mu \, b] [\bar \ell \, \gamma_\mu \, P_L \, \nu]  \nonumber \\ 
{\cal O}^{cb\ell}_{\rm AL}       &=    [\bar{c} \, \gamma^\mu \, \gamma_5 \, b] [\bar \ell \, \gamma_\mu \, P_L \, \nu] \nonumber \\  
{\cal O}^{cb\ell}_{\rm SL}       &=    [\bar{c} \, b] [\bar \ell  \, P_L \, \nu] \nonumber \\ 
{\cal O}^{cb\ell}_{\rm PL}       &=    [\bar{c} \, \gamma_5 \, b] [[\bar \ell \, P_L \, \nu] \nonumber \\ 
{\cal O}^{cb\ell}_{\rm TL}       &=   [\hedt] [\bar \ell \, \sigma_{\mu\nu} \, P_L \, \nu] \nonumber 
\end{align}
\begin{align}
{\cal O}^{cb \ell}_{\rm VR}      &=    [\bar{c} \, \gamma^\mu \, b] [\bar \ell \, \gamma_\mu \, P_R \, \nu]  \nonumber \\ 
{\cal O}^{cb\ell}_{\rm AR}       &=    [\bar{c} \, \gamma^\mu \, \gamma_5 \, b] [\bar \ell \, \gamma_\mu \, P_R \, \nu] \nonumber \\  
{\cal O}^{cb\ell}_{\rm SR}       &=    [\bar{c} \, b] [\bar \ell  \, P_R \, \nu] \\ 
{\cal O}^{cb\ell}_{\rm PR}       &=    [\bar{c} \, \gamma_5 \, b] [[\bar \ell \, P_R \, \nu] \nonumber \\ 
{\cal O}^{cb\ell}_{\rm TR}       &=   [\hedt] [\bar \ell \, \sigma_{\mu\nu} \, P_R \, \nu] \nonumber 
\end{align}
\end{multicols} \vspace{-4mm}
and the set of Wilson Coeffients (WCs) corresponding to these operators are defined at the renormalization scale $\mu = m_b$.

In the SM, we would have $C^{cb\ell}_{\rm VL} = - C^{cb\ell}_{\rm AL} = 1$. We wish to go beyond 
the SM, but we shall respect the full gauge invariance of the SM and consequently only consider the operators listed on the left in \ref{b2c-basis}.
Further, since it is difficult to build a microscopic model with tensor interaction, we neglect its contribution in this note.(For the study of tensor operators, refer to the Appendix of \cite{Bardhan:2016uhr}).
\vspace{-5mm}
\section{Form Factors}
\subsection{For $B \to D$ decay} 
The non-zero hadronic matrix elements for $\bar{B} \to D$ transition (ignoring the tensor) are parameterized 
by
\bea\label{b2d_ff}
\bra D(p_D, M_D)| \bar c \gamma^\mu b |\bar B(p_B, M_B)\ket &=& F_+(q^2) \Big[(p_B+p_D)^\mu -\frac{M_B^2 - M_D^2}{q^2} q^\mu \Big] \nonumber \\*
&&  + F_0(q^2) \frac{M_B^2- M_D^2}{q^2} q^\mu \nn \\*
\bra D(p_D, M_D)| \bar c b |\bar B(p_B, M_B)\ket &=& F_0(q^2) \frac{M_B^2 -  
M_D^2}{m_b - m_c} 
\eea
Calculations for the form factors $F_0(q^2)$ and $F_+(q^2)$ are known in a lattice framework \cite{Bailey:2014tva}. 
The axial vector and the pseudoscalar matrix elements are zero from symmetry considerations
and thus only the WCs $C_{\rm VL}^\tau$ and $C_{\rm SL}^\tau$ contribute to this decay. 
\subsection{For $B \to D^\ast$ decay} 
The non-zero hadronic matrix elements for $\bar{B} \to D^*$ transition are parametrised by
\bea
\label{b2ds_ff}
\bra D^\ast(p_{D^\ast}, M_{D^\ast})| \bar c \gamma_\mu b |\bar B(p_B, M_B)\ket &=&  
i \varepsilon_{\mu\nu\rho\sigma} \epsilon^{ \nu \ast} p_B^\rho p_{D^\ast}^\sigma  \, \frac{2 V(q^2)}{M_B+M_{D^\ast}} \nn\\
\bra D^\ast(p_{D^\ast}, M_{D^\ast})| \bar c \gamma_\mu \gamma_5 b |\bar B(p_B,
M_B)\ket &=&  2 M_{D^\ast}  \frac{\epsilon^\ast.q }{q^2 } q_\mu A_0(q^2) +
(M_B+M_{D^\ast})  \Big[\epsilon_{\mu}^\ast - \frac{\epsilon^\ast.q}{q^2 } q_\mu   \Big] A_1(q^2) \nn\\*
&& \hspace*{-2mm} - \frac{\epsilon^\ast.q}{M_B+M_{D^\ast}} \Big[(p_B + p_{D^\ast})_\mu  -\frac{M_B^2-M_{D^\ast}^2}{q^2} q_\mu  \Big] A_2(q^2) \nn\\
\bra D^\ast(p_{D^\ast}, M_{D^\ast})| \bar c \gamma_5 b |\bar B(p_B, M_B)\ket &=&  
- \epsilon^\ast.q \, \frac{2 M_{D^\ast}}{m_b + m_c}  \, A_0(q^2) 
\eea
While no lattice calculations exist for the form factors in this case, they have been calculated in a Heavy Quark Effective
Theory (HQET) framework \cite{Caprini:1997mu} and we borrow those results. In this case, symmetry dictates that the scalar current is zero and thus
there is no contribution to the decay width from $C_{SL}^\tau$.
\subsection{Independence of $R_D$ and $R_{D^\ast}$} 
We see that while $C_{VL}^\tau$ and $C_{SL}^\tau$ contribute to the $B\to D$ decay process, $C_{VL}^\tau$, $C_{AL}^\tau$ and
$C_{PL}^\tau$ contribute to the other one. Thus, given the independence of the WCs, the two processes are independent of each
other since they depend of different sets of WCs. In other words, $\rd$ and $\rds$ are theoretically independent measurements 
and allow for separate explanations. 
\section{Explaining $R_D$ Alone}
\begin{figure}[h!]
\begin{center}
\begin{tabular}{cc}
\includegraphics[scale=0.3]{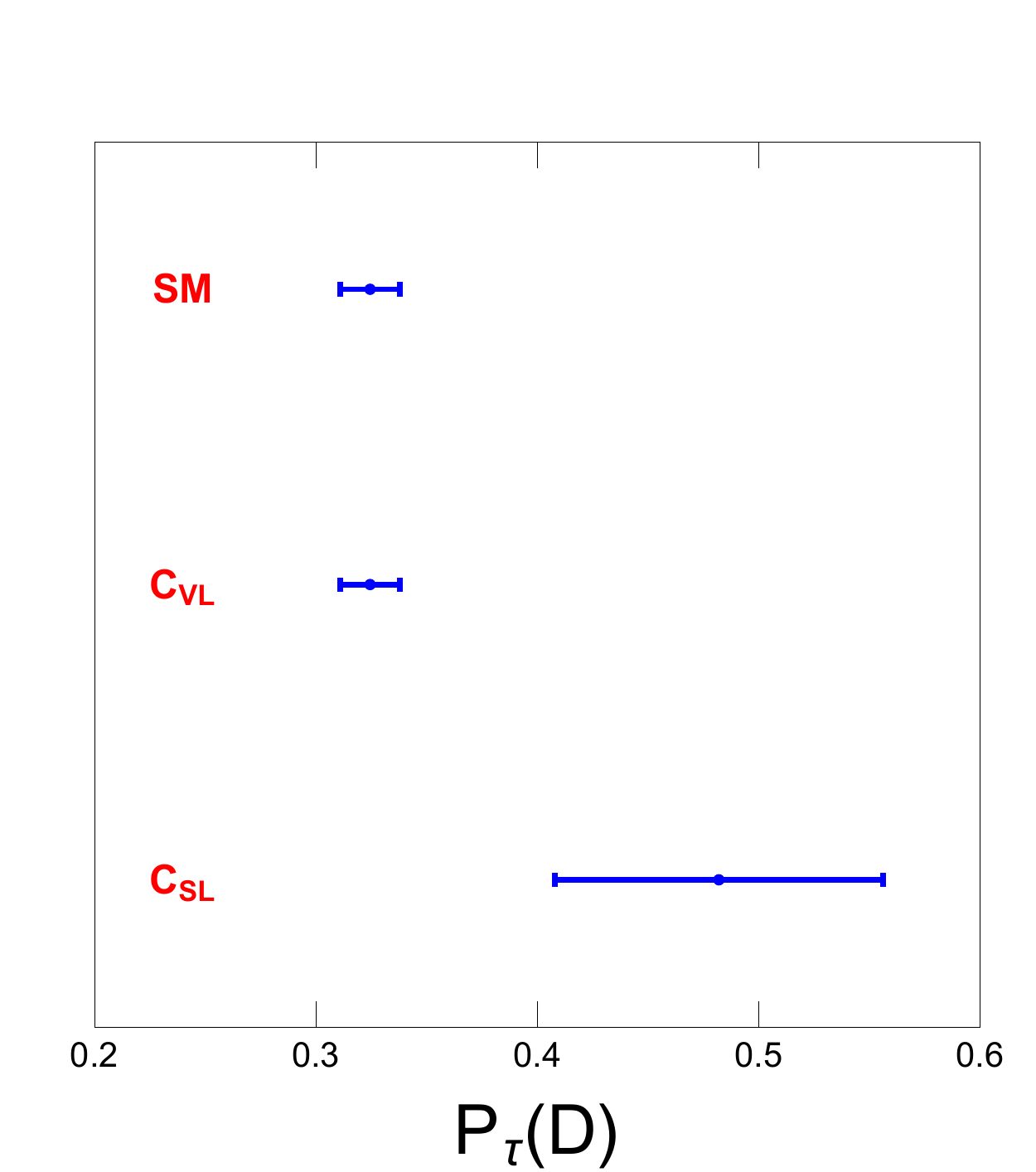}
\includegraphics[scale=0.3]{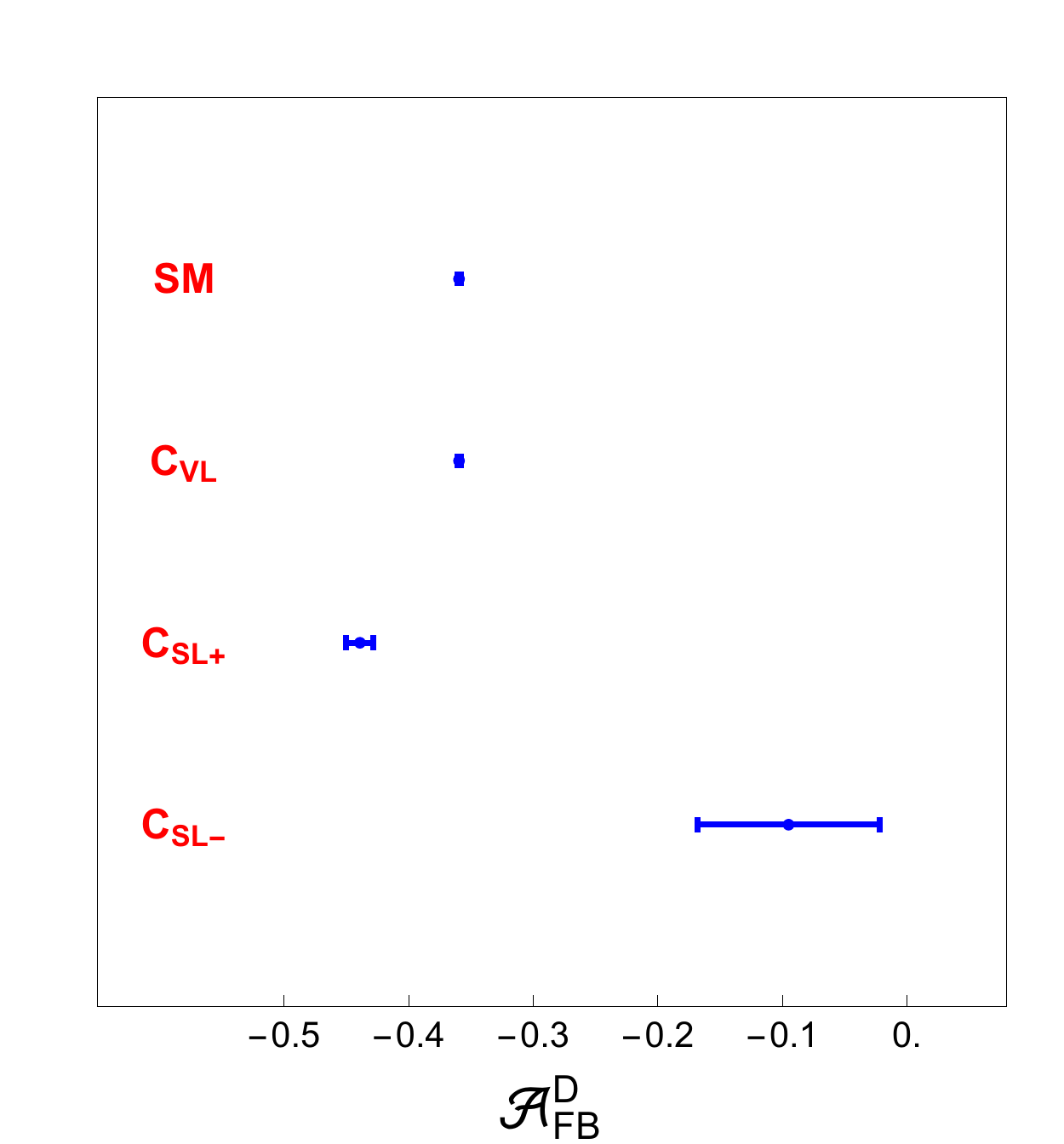}
\end{tabular}
\caption{Predictions for the polarisation fraction $P_\tau(D)$ (left) and $\mathcal{A}_{FB}^D$ (right)
\label{fig:ptau-fbad}}
\end{center}
\end{figure}

The quantities $a_\ell^D$, $b_\ell^D$ and $c_\ell^D$ (in \ref{eqn:br}) can be calculated for a particular helicity of the final state lepton 
using a helicity amplitude approach. The complete expressions are given in \cite{Bardhan:2016uhr} and it is not repeated here. Since only
$C_{VL}^\tau$ and $C_{SL}^\tau$ are relevant, we can plot $\rd$ as a variation of the two WCs and note the range of values for which
it satisfies the experimental bounds. 
\begin{figure}[h!]
\centering
\includegraphics[scale=0.3]{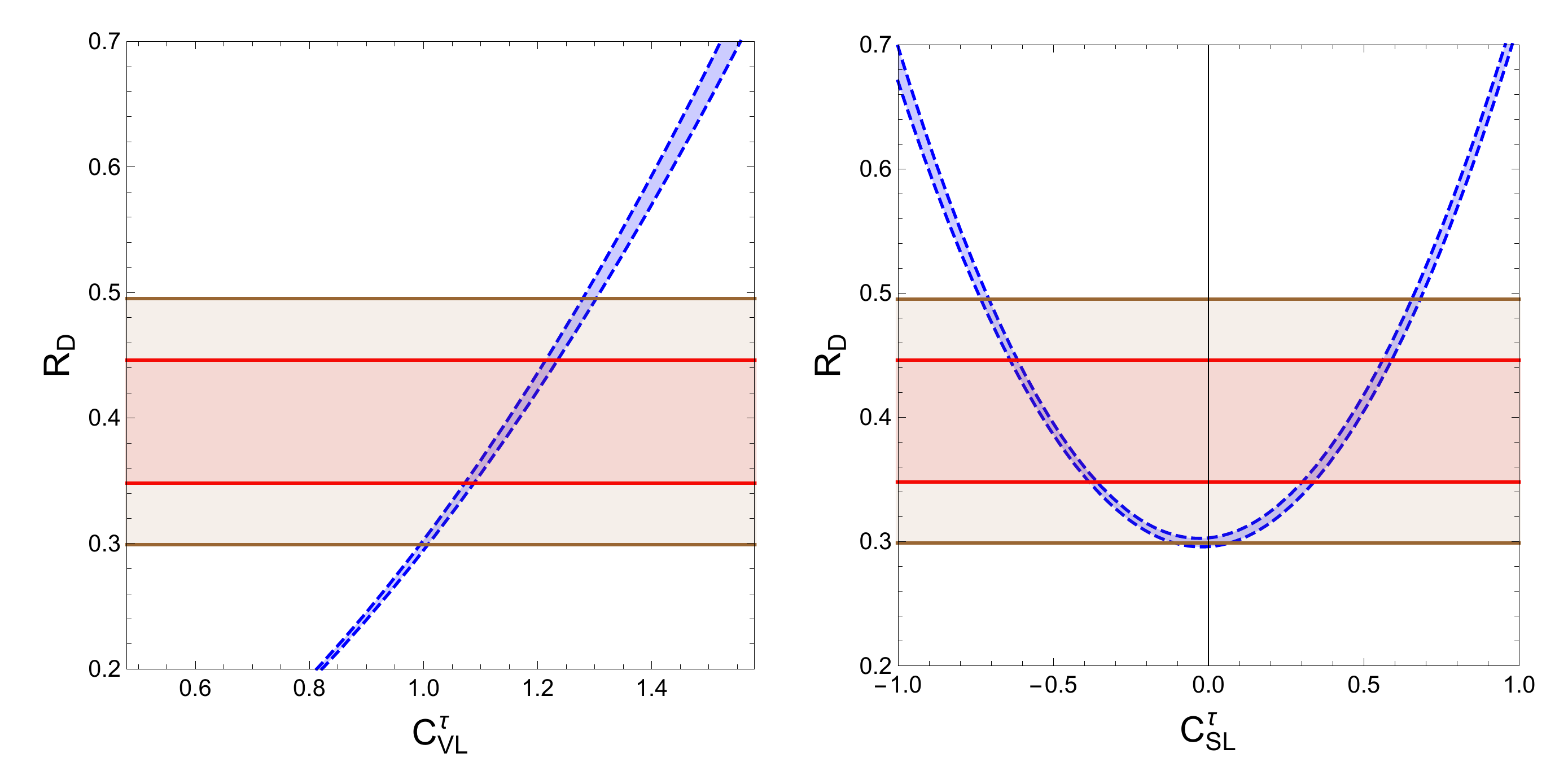}
\caption{The dependence of $R_D$ with respect to the variation of the WCs $C_{VL}^\tau$ (left) and 
$C_{SL}^\tau$ (right). \label{rd-vs-wc}}
\end{figure}
This is done in Fig. \ref{rd-vs-wc}, where the red (brown) band corresponds to the $1 \sigma$ ($2\sigma$) 
value on the experimental measurement. 
\begin{figure}[h!]
\begin{center}
\includegraphics[scale=0.3]{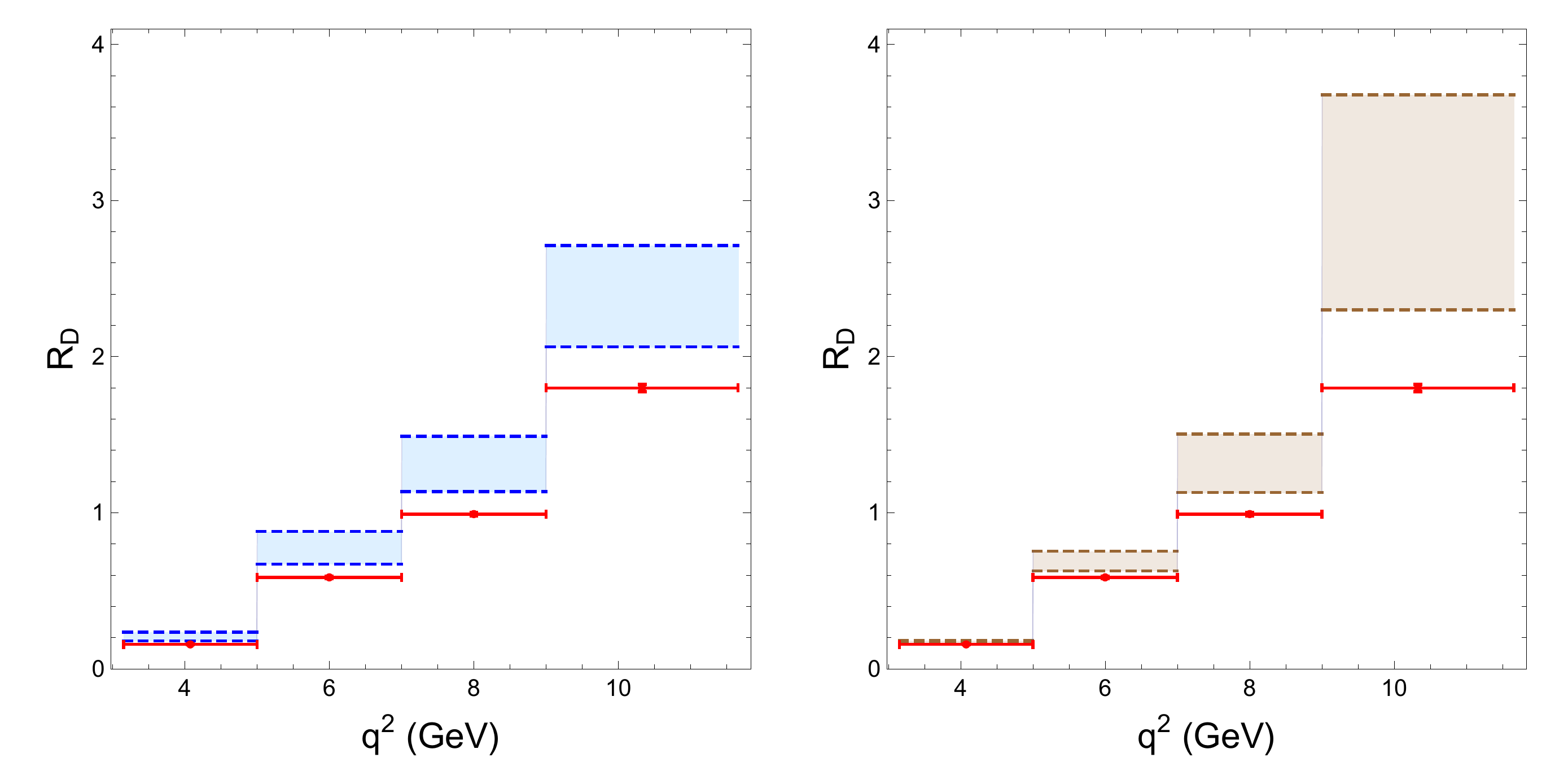}
\caption{The binwise $R_D$ for four $q^2$ bins. On the left, $C_{VL}^\tau$ is varied, while on the right, 
$C_{SL}^\tau$ is varied within their $1 \sigma$ allowed ranges.\label{fig:binned-rd}}
\end{center}
\end{figure}
We can use this range of the WCs to make a prediction of the value of the binned $\rd$, and for
the values of $P_\tau^D$ and $\fbad$. These are shown in Fig. \ref{fig:binned-rd} and Fig. \ref{fig:ptau-fbad} respectively. 

\section{Explaining $R_{D^\ast}$ Alone}
\begin{figure}[!h]
\begin{center}
\begin{tabular}{l}
\includegraphics[scale=0.35]{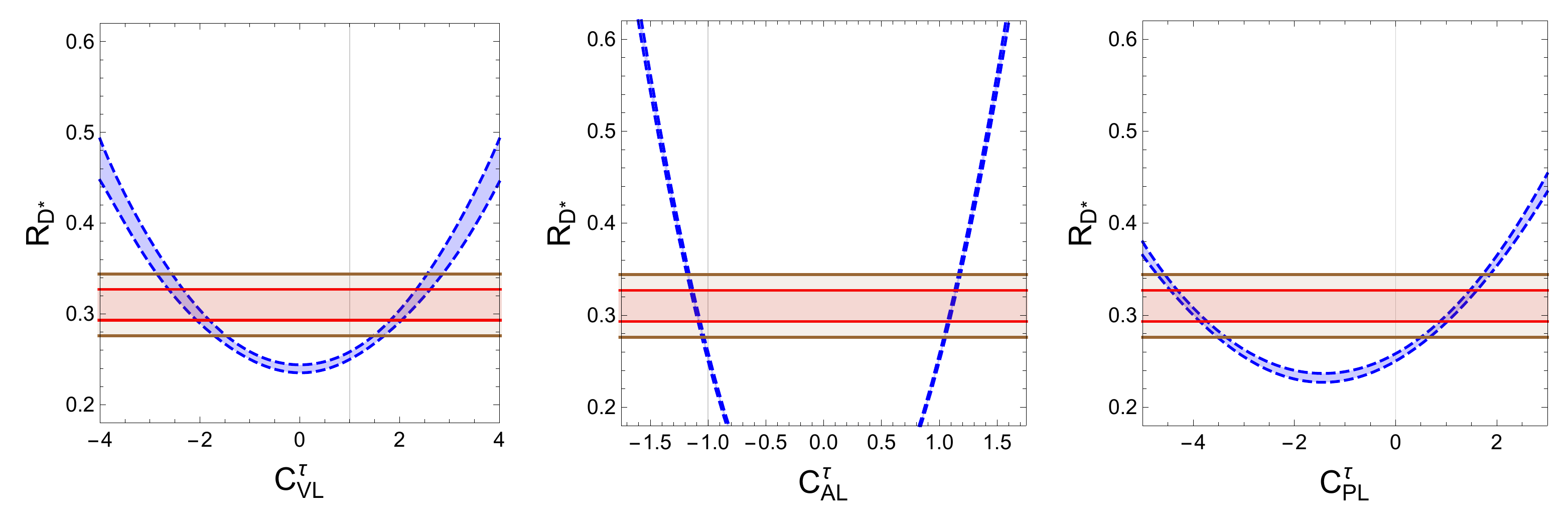} 
\end{tabular}
\caption[]{The dependence of $R_{D^*}$ with respect to the variation of the WCs  
$C_{VL}^\tau$ (left), $C_{AL}^\tau$ (middle) and $C_{PL}^\tau$ (right). A thin vertical line shows the SM values of the WCs.
\label{rds-vs-wc}}
\end{center} 
\end{figure}

We can carry out a similar treatment for the case of $B \to D^\ast$ decay. In this case, three WCs - $C_{VL}^\tau$, $C_{AL}^\tau$ and
$C_{PL}^\tau$ - contribute. The plots of $\rds$ as a function of the different WCs are given in Fig. \ref{rds-vs-wc} As before, the $1\sigma$ ($2\sigma$)
bands are indicated by the red (brown) bands. 
\begin{figure}[h!]
\begin{center}
\includegraphics[scale=0.35]{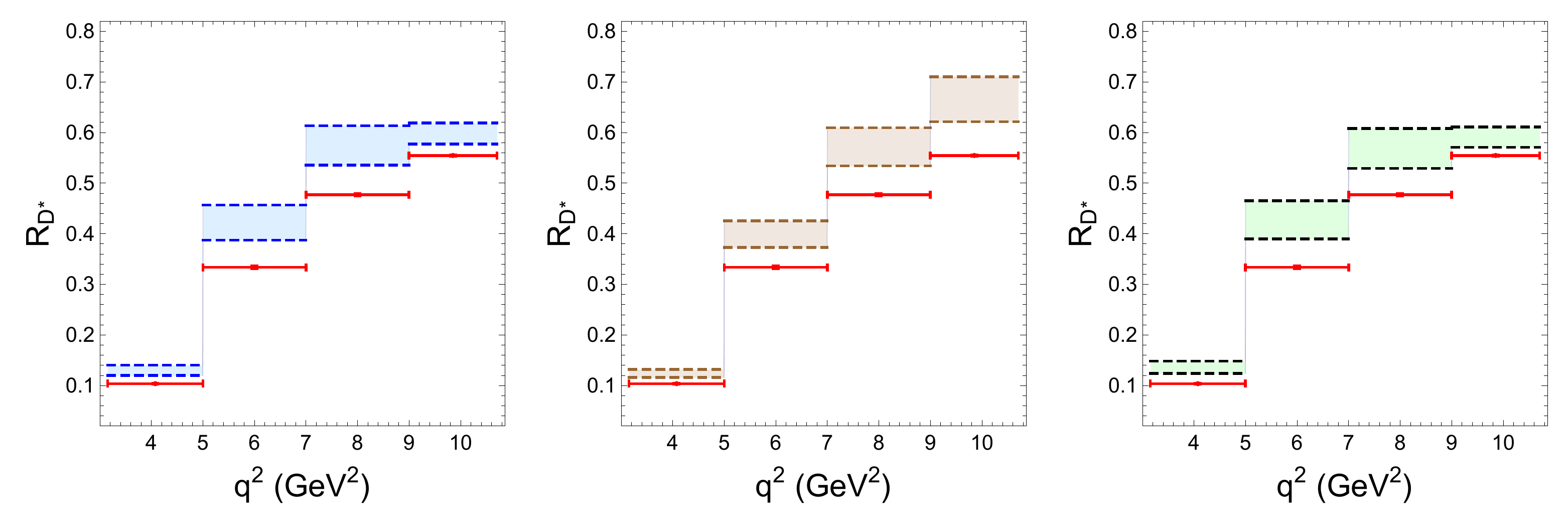}
\caption{The binwise $R_D^*$ for four $q^2$ bins. On the left, $C_{VL}^\tau$ is varied, in the middle $C_{AL}^\tau$ 
is varied, annd on the right, $C_{PL}^\tau$ is varied within their $1 \sigma$ allowed ranges. The SM predictions are shown in red. 
\label{fig:binned-rds}}
\end{center}
\end{figure}
\begin{figure}[h!]
\begin{center}
\begin{tabular}{cc}
\includegraphics[scale=0.31]{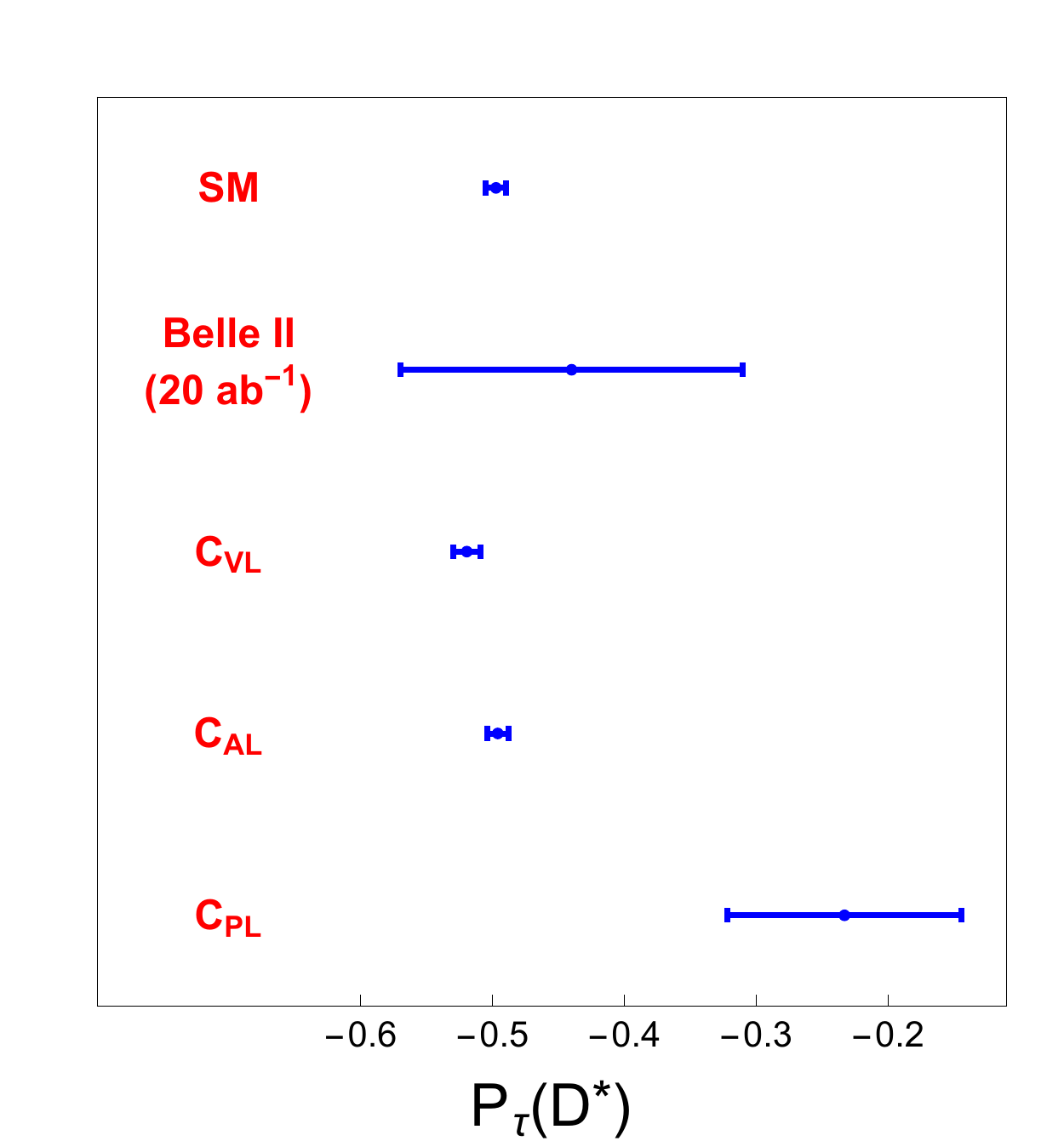}
\includegraphics[scale=0.3]{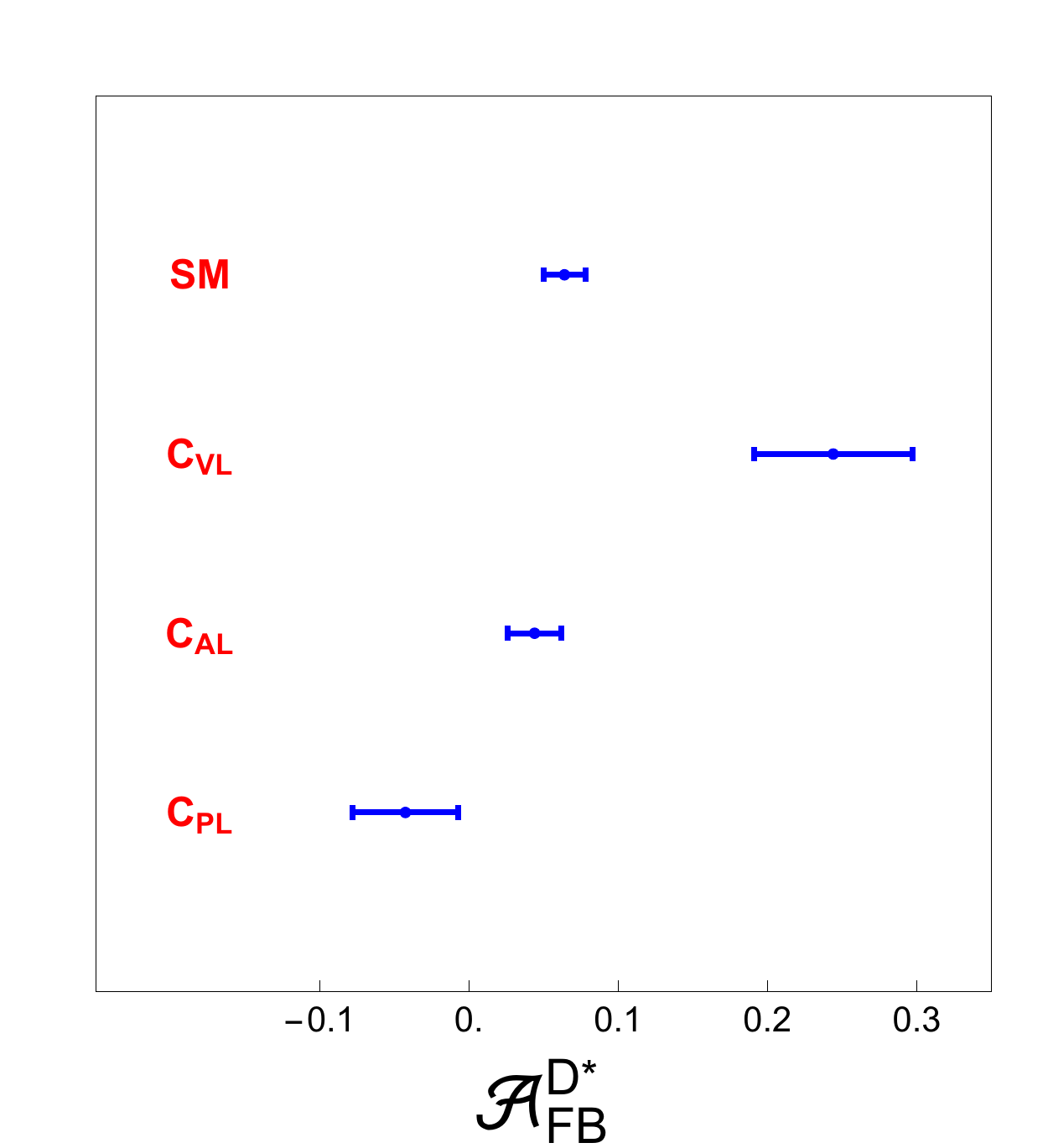}
\end{tabular}
\caption{Predictions for the polarisation fraction $P_\tau(D^*)$ (left), $\mathcal{A}_{FB}^{D^\ast}$ (right). In the left plot, the Belle II 20 ab$^{-1}$ 
projection is shown.}
\label{fig:ptau-fbads}
\end{center}
\end{figure}
The prediction for the binned $\rds$ is given in Fig. \ref{fig:binned-rds}. In this case, we do have a 
measurement of $P_\tau^{D^\ast}$, but it is quite imprecise. In the left plot of Fig. \ref{fig:ptau-fbads}, the size of the errors indicated for 
the {\sc Belle} measurement is a projection with 20 $ab^{-1}$ data, which is expected to be collected by the year 2021; the central value
indicated is the current central value. As a matter of completion, we also plot the prediction for $\fbads$ on the right of Fig. \ref{fig:ptau-fbads}, 
although no measurement of this quantity exists as yet. 
\begin{figure}[h!]
\begin{center}
\begin{tabular}{c}
\includegraphics[scale=0.32]{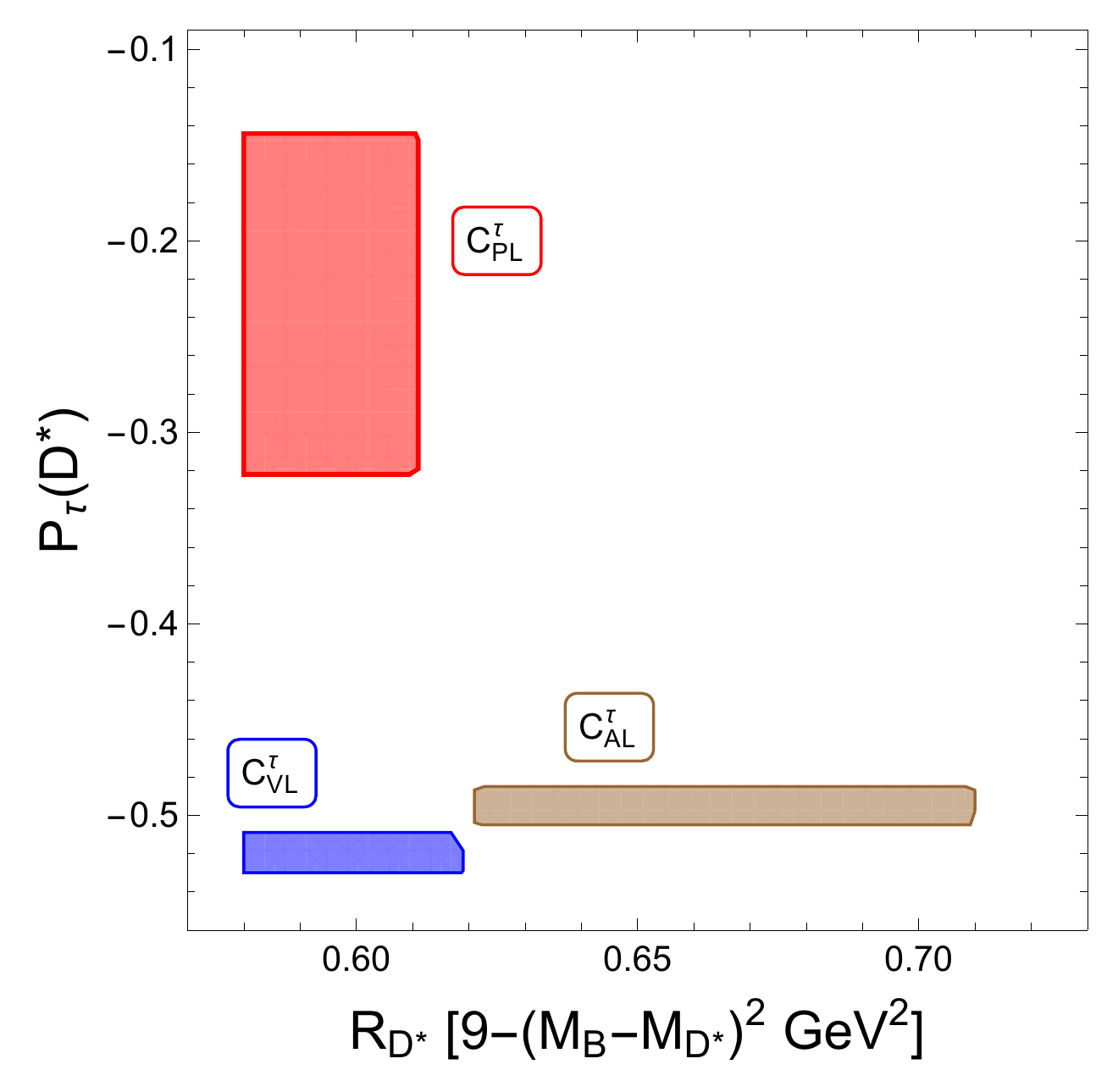}
\includegraphics[scale=0.32]{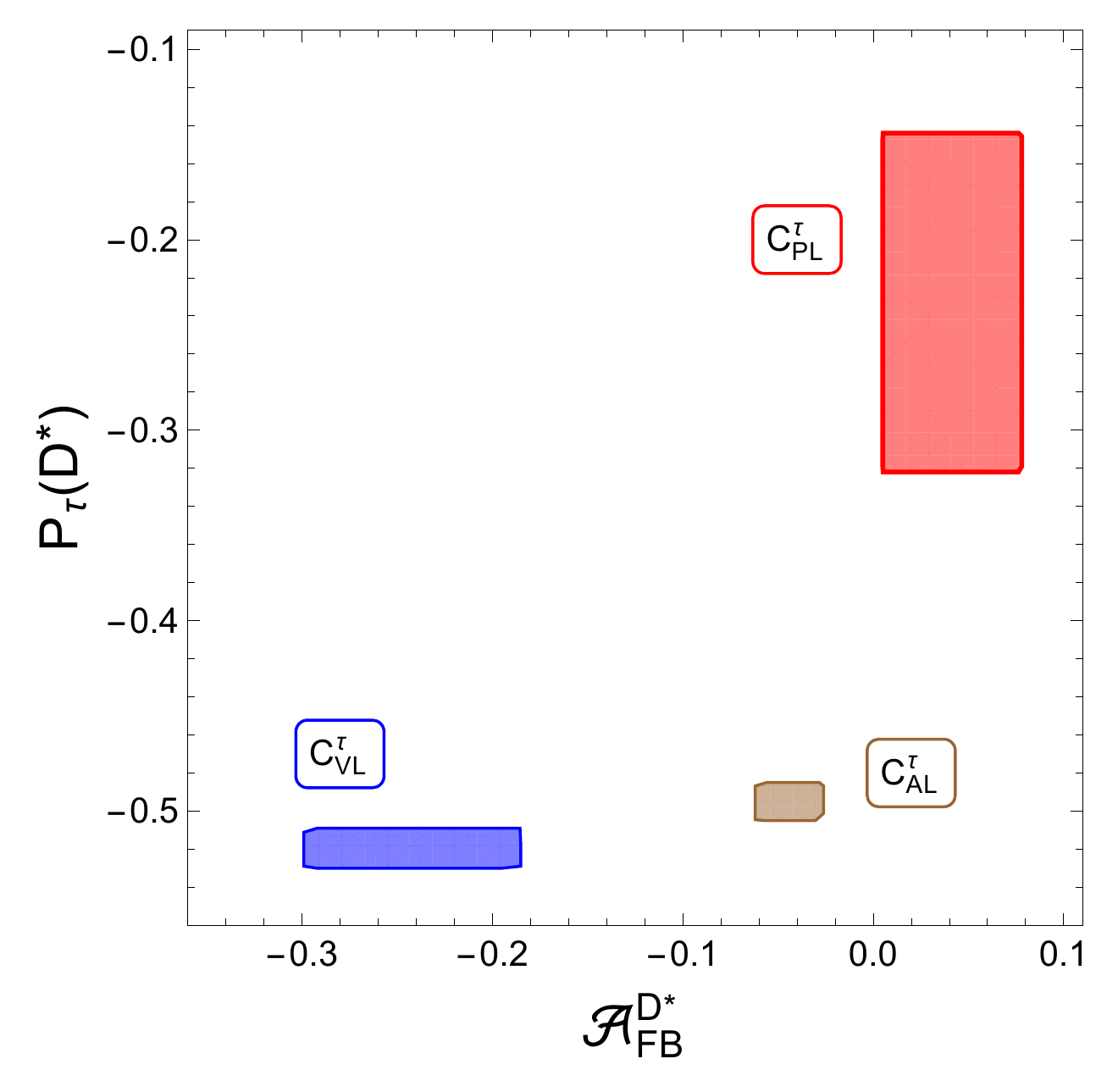}
\includegraphics[scale=0.32]{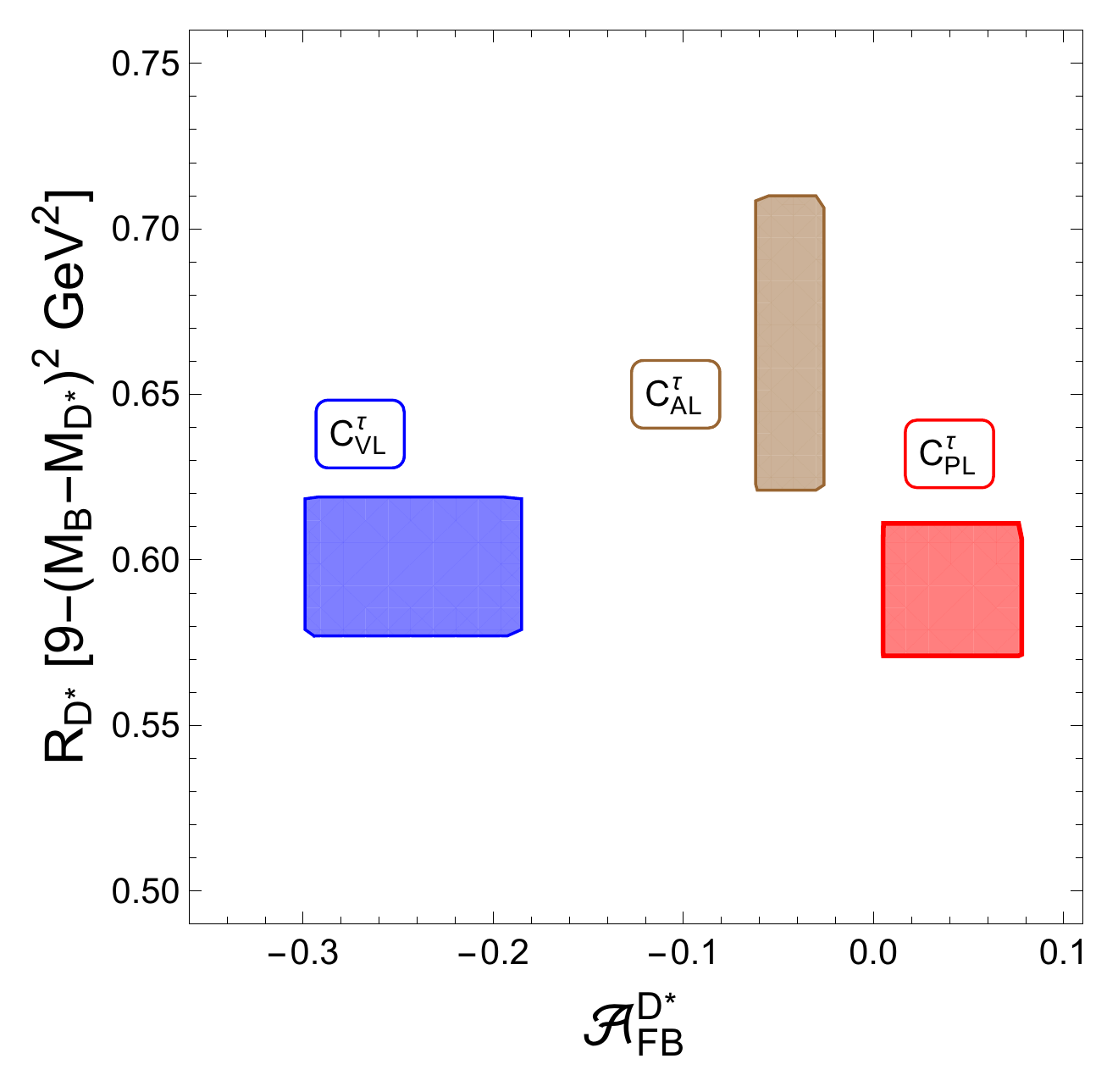}
\end{tabular}
\caption{ The predictions for $P_\tau^{D^\ast}$, $R_{D^*}$ in the last bin and $\mathcal{A}_{FB}^{D^\ast}$ are shown in three different planes
 for the ranges of the three WCs $C_{VL}^\tau$, $C_{AL}^\tau$ and $C_{PL}^\tau$. 
 \label{fig:RDs-Ptau-fbads}}
\end{center}
\end{figure}
We can combine the predictions for the binned $\rds$ restricted to the highest $q^2$ bin, $P_\tau^{D^\ast}$ and $\fbads$ to construct three planes. When
plotted in these three planes, the regions of the allowed values of the WCs all separate out nicely as shown in Fig. \ref{fig:RDs-Ptau-fbads}. A future measurement of any two of these three
observables would help in restricting us to a particular region, thus limiting the scope of any NP model.

\end{document}